\documentclass[10pt,twocolumn,letterpaper]{article}

\usepackage{cvpr}
\usepackage{times}
\usepackage{epsfig}
\usepackage{graphicx}
\usepackage{amsmath}
\usepackage{amssymb}
\usepackage{enumitem}
\usepackage{soul}
\usepackage{multirow, booktabs}
\usepackage{subcaption}
\usepackage{balance}
\usepackage{xcolor}
\usepackage{makecell}
\usepackage{caption}
\usepackage{fancyhdr}

\usepackage[pagebackref=true,breaklinks=true,letterpaper=true,colorlinks,bookmarks=false]{hyperref}

\usepackage{cleveref}
\cvprfinalcopy 

\def\cvprPaperID{24} 

\ifcvprfinal\fi
\begin{document}

\title{Representation Learning of Histopathology Images using Graph Neural Networks}

\author{Mohammed Adnan$^{1,2,}\thanks{Authors have contributed equally.}$, Shivam Kalra$^{1,*}$, Hamid R. Tizhoosh$^{1,2}$\\
$^1$Kimia Lab, University of Waterloo, Canada\\
$^2$Vector Institute, Canada\\
{\small\tt\{m7adnan,shivam.kalra,tizhoosh\}@uwaterloo.ca}
}

\maketitle
\begin{abstract}
Representation learning for Whole Slide Images (WSIs) is pivotal in developing
image-based systems to achieve higher precision in diagnostic pathology. We
propose a two-stage framework for WSI representation learning. We sample
relevant patches using a color-based method and use graph neural networks to
learn relations among sampled patches to aggregate the image information into a
single vector representation. We introduce attention via graph pooling to
automatically infer patches with higher relevance. We demonstrate the
performance of our approach for discriminating two sub-types of lung cancers,
Lung Adenocarcinoma (LUAD) \& Lung Squamous Cell Carcinoma (LUSC). We collected
1,026 lung cancer WSIs with the 40$\times$ magnification from The Cancer Genome
Atlas~(TCGA) dataset, the largest public repository of histopathology images and
achieved state-of-the-art accuracy of 88.8\% and AUC of 0.89 on lung cancer
sub-type classification by extracting features from a pre-trained DenseNet
model.
\end{abstract}

\section{Introduction}
Large archives of digital scans in pathology are gradually becoming a reality.
The amount of information stored in such archives is both impressive and
overwhelming. However, there is no convenient provision to access this stored
knowledge and make it available to pathologists for diagnostic, research, and
educational purposes. This limitation is mainly due to the lack of techniques
for representing WSIs. The characterization of WSIs offers
various challenges in terms of image size, complexity and color, and
definitiveness of diagnosis at the pathology level, also the sheer amount of
effort required to annotate a large number of images. These challenges
necessitate inquiry into more effective ways of representing WSIs.

The recent success of ``deep learning'' has opened promising horizons for
digital pathology. This has motivated both AI experts and pathologists to work
together in order to create novel diagnostic algorithms. This opportunity has
become possible with the widespread adoption of \emph{digital pathology}, which
has increased the demand for effective and efficient analysis of Whole Slide
Images (WSIs). Deep learning is at the forefront of computer vision, showcasing
significant improvements over conventional methodologies on visual
understanding. However, each WSI consist of billions of pixel and therefore,
deep neural networks cannot process them. Most of the recent work analyzes WSIs at the
patch level, which requires manual delineation from an expert. Therefore, the
feasibility of such approaches is rather limited for larger archives of WSIs.
Moreover, most of the time, labels are available for the entire WSI, and not for
individual patches. To learn a representation of a WSI, it is, therefore,
necessary to leverage the information present in all patches. Hence, multiple
instance learning (MIL) is a promising approach for learning WSI representation.

MIL is a type of supervised learning approach which uses a set of instances
known as a bag. Each bag has an associated label instead of individual
instances. MIL is thus a natural candidate for learning WSI representation. We
explore the application of graph neural networks for MIL. We propose a framework
that models a WSI as a fully connected graph to extract its representation. The
proposed method processes the entire WSI at the highest magnification level; it
requires a single label of the WSI without any patch-level annotations.
Furthermore, modeling WSIs as fully-connected graphs enhance the
interpretability of the final representation. We treat each instance as a node
of the graph to learn relations among nodes in an end-to-end fashion. Thus, our
proposed method not only learns the representation for a given WSI but also
models relations among different patches. We explore our method for classifying
two subtypes of lung cancer, Lung Adenocarcinoma (LUAD) and Lung Squamous Cell
Carcinoma (LUSC). LUAD and LUSC are the most prevalent subtypes of lung cancer,
and their distinction requires visual inspection by an experienced pathologist.
In this study, we used WSIs from the largest publicly available dataset, The
Cancer Genome Atlas (TCGA)~\cite{weinstein2013cancer}, to train the model for
lung cancer subtype classification. We propose a novel architecture using graph
neural networks for learning WSI representation by modeling the relation among
different patches in the form of the adjacency matrix. Our proposed method
achieved an accuracy of 89\% and 0.93 AUC. The contributions of the paper are
3-folds, i) a graph-based method for representation learning of WSIs, and ii) a
novel adjacency learning layer for learning connections within nodes in an
end-to-end manner, and iii) visualizing patches which are given higher
importance by the network for the prediction.

The paper is structured as follows: Section~\ref{sec:relwork} briefly covers the
related work. Section~\ref{sec:background} discusses Graph Convolution Neural
Networks (GCNNs) and deep sets. Section~\ref{sec:app} explains the approach, and
experiments \& results are reported in Section~\ref{sec:result}.

\section{Related Work}
\label{sec:relwork}
With an increase in the workload of pathologists, there is a clear need to
integrate CAD systems into pathology
routines~\cite{KomuraMachinelearningmethods2017,
MadabhushiImageanalysismachine2016,
MadabhushiComputeraidedprognosisPredicting2011,
GurcanHistopathologicalImageAnalysis2009}. Researchers in both image analysis
and pathology fields have recognized the importance of quantitative image analysis
 by using machine learning (ML)
techniques~\cite{GurcanHistopathologicalImageAnalysis2009}. The continuous
advancement of digital pathology scanners and their proliferation in clinics and
laboratories have resulted in a substantial accumulation of histopathology
images, justifying the increased demand for their analysis to improve the
current state of diagnostic pathology~\cite{MadabhushiImageanalysismachine2016,
KomuraMachinelearningmethods2017}. \\

\noindent \textbf{Histopathology Image Representation.} To develop CAD systems
for histopathology images (WSIs), it is crucial to transform WSIs into feature
vectors that capture their diagnostic semantics. There are two main methods for
characterizing WSIs~\cite{barker2016automated}. The first method is called
\emph{sub-setting method}, which considers a small section of a large pathology
image as an essential part such that the processing of a small subset
substantially reduces processing time. A majority of research studies in the
literature have used the \emph{sub-setting} method because of its speed and
accuracy. However, this method requires expert knowledge and intervention to extract the
proper subset. On the other hand, the \emph{tiling method} segments images into
smaller and controllable patches (i.e., tiles) and tries to process them against
each other~\cite{gutman2013cancer}, which requires a more automated approach.
The tiling method can benefit from MIL; for example, Ilse et
al.~\cite{ilse2018attention} used MIL with attention to classify breast and
colon cancer images.

Due to the recent success of artificial intelligence (AI) in computer vision
applications, many researchers and physicians expect that AI would be able to
assist physicians in many tasks in digital pathology. However, digital pathology images are
 difficult to use for training neural networks. A single WSI
is a gigapixel file and exhibits high
morphological heterogeneity and may as well as contain different  artifacts.
All in all, this impedes the common use of deep
learning~\cite{dimitriou2019deep}. \\

\noindent \textbf{Multiple Instance Learning (MIL).} MIL algorithms assign a
class label to a set of instances rather than to individual instances. The
individual instance labels are not necessarily important, depending on the type
of algorithm and its underlying assumptions. \cite{carbonneau2018multiple}.
Learning representation for histopathology images can be formulated as a MIL
problem. Due to the intrinsic ambiguity and difficulty in obtaining human
labeling, MIL approaches have their particular advantages in automatically
exploiting the fine-grained information and reducing efforts of human
annotations. Isle et al. used MIL for digital pathology and introduces a
different variety of MIL pooling functions~\cite{ilse2020deep}. Sudarshan et al.
used MIL for histopathological breast cancer image
classification~\cite{sudharshan2019multiple}. Permutation invariant operator for
MIL was introduced by Tomczak et al. and successfully applied to digital
pathology images~\cite{tomczak2017deep}. Graph neural networks (GNNs) have been
used for MIL applications because of their permutation invariant
characteristics. Tu et al. showed that GNNs can be used for MIL, where each
instance acts as a node in a graph~\cite{tu2019multiple}. Anand et al. proposed
a GNN based approach to classify WSIs represented by graphs of its constituent
cells~\cite{anand2020histographs}.

\section{Background}
\label{sec:background}
\noindent \textbf{Graph Representation.} A graph can be fully represented by its
node list $V$ and adjacency matrix $\mathbf{A}$. Graphs can model many types of
relations and processes in physical, biological, social, and information systems.
A connection between two nodes $V_i$ and $V_j$ is represented using an edge weighted by $a_{ij}$.\\

\noindent \textbf{Graph Convolution Neural Networks (GCNNs).} GCNNs generalize
the operation of convolution from grid data to graph data. A GCNN takes a graph
as an input and transforms it into another graph as the output. Each feature
node in the output graph is computed by aggregating features of the
corresponding nodes and their neighboring nodes in the input graph. Like CNNs,
GCNNs can stack multiple layers to extract high-level node representations.
Depending upon the method for aggregating features, GCNNs can be divided into
two categories, namely spectral-based and spatial-based. Spectral-based
approaches define graph convolutions by introducing filters from the perspective
of graph signal processing. Spectral convolutions are defined as the
multiplication of a node signal by a kernel. This is similar to the way
convolutions operate on an image, where a pixel value is multiplied by a kernel
value. Spatial-based approaches formulate graph convolutions as aggregating
feature information from neighbors. Spatial graph convolution learns the
aggregation function, which is permutation invariant to the ordering of the
node.\\

\noindent \textbf{ChebNet.} It was introduced by Defferrard et
al.~\cite{defferrard2016convolutional}. Spectral convolutions on graphs are defined
as the multiplication of a signal $x \in R^N$ (a scalar for every node) with a
filter $g(\theta) = diag(\theta)$ parameterized by $\theta \in R^N$ in the
Fourier domain, i.e.,
\begin{equation*}
    g_{\theta}   \circledast x = U g_{\theta}U^{T}x,
\end{equation*}
where $U$ is the matrix of eigenvectors of the normalized graph Laplacian $L =
I_{N} - D^{-\frac{1}{2}}AD^{-\frac{1}{2}}$. This equation is computationally
expensive to calculate as multiplication with the eigenvector matrix U is
$O(N^2)$. Hammond et al.~\cite{hammond2011wavelets} suggested that that
$g_{\theta}$ can be well-approximated by a truncated expansion in terms of
Chebyshev polynomials $T_k(x)$, i.e,
\begin{equation*}
    g_{\theta^\prime} (\Lambda) \approx \sum_{k=0}^{K} \theta^{\prime}T_{k}(\Lambda).
\end{equation*}

\begin{figure*}[htb]
    \centering    
    \includegraphics[width=\textwidth]{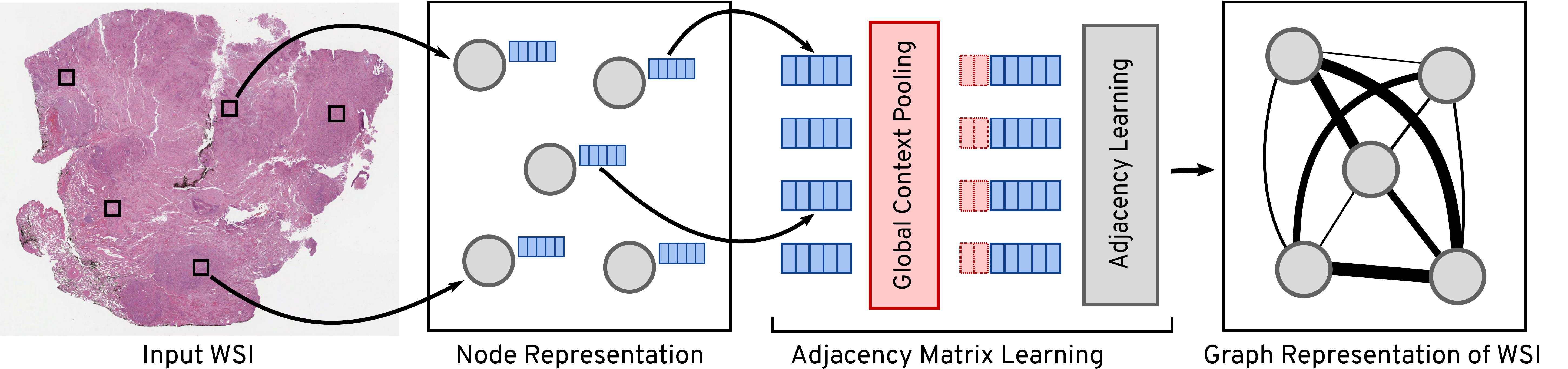}
    \vspace{0.1em}
    \caption{\textbf{Transforming a WSI to a fully-connected graph.} A WSI is
      represented as a graph with its nodes corresponding to distinct patches from the
      WSI. A node feature (a blue block beside each node) is extracted by feeding the
      associated patch through a deep network. A single context vector, summarizing
      the entire graph is computed by pooling all the node features. The context
      vector is concatenated with each node feature, subsequently fed into adjacent
      learning block. The adjacent learning block uses a series of dense layers and
      cross-correlation to calculate the adjacency matrix. The computed adjacency
      matrix is used to produce the final fully-connected graph. In the figure,
      the thickness of the edge connecting two nodes corresponds to the value in the
      adjacency matrix.}
    \label{fig:wsi_graph}
  \end{figure*}

The kernels used in ChebNet are made of Chebyshev polynomials of the diagonal
matrix of Laplacian eigenvalues. ChebNet uses kernel made of Chebyshev
polynomials. Chebyshev polynomials are a type of orthogonal polynomials with
properties that make them very good at tasks like approximating functions.\\

\noindent \textbf{GraphSAGE.} It was introduced by Hamilton et
al.~\cite{hamilton2017inductive}. GraphSAGE learns aggregation functions that
can induce the embedding of a new node given its features and neighborhood. This
is called inductive learning. GraphSAGE is a framework for inductive
representation learning on large graphs that can generate low-dimensional vector
representations for nodes and is especially useful for graphs that have rich
node attribute information. It is much faster to create embeddings for new nodes
with GraphSAGE.\\

\noindent \textbf{Graph Pooling Layers.} Similar to CNNs, pooling
layers in GNNs downsample node features by pooling operation.
We experimented with Global Attention Pooling, Mean Pooling, Max Pooling, and
Sum Pooling. Global Attention Pooling~\cite{li2015gated} was introduced by Li et
al. and uses soft attention mechanism to decide which nodes are relevant to the
current graph-level task and gives the pooled feature vector from all the
nodes.\\

\noindent \textbf{Universal Approximator for Sets.} We use results from Deep
Sets~\cite{zaheer2017deep} to get the global context of the set of patches
representing WSI. Zaheer et al. proved in~\cite{zaheer2017deep} that any set can
be approximated by $\rho \sum(\phi(x))$ where $\rho$ and $\phi$ are some
function, and $x$ is the element in the set to be approximated.

\section{Method}
\label{sec:app}
\begin{figure*}[h]
  \centering
  \includegraphics[width=0.85\textwidth]{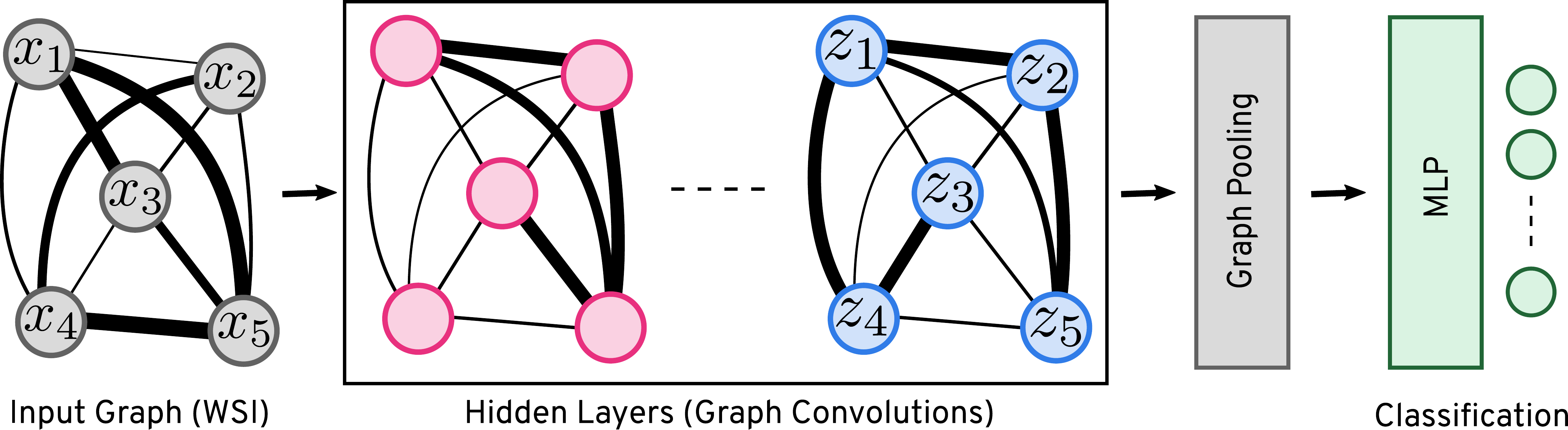}
  \caption{\textbf{Classification of a graph representing a WSI.} A fully
    connected graph representing a WSI is fed through a graph convolution layer to
    transform it into another fully-connected graph. After a series of
    transformations, the nodes of the final fully-connected graph are aggregated
    to a single condensed vector, which is fed to an MLP for classification purposes. }
  \label{fig:my_label}
\end{figure*}

The proposed method for representing a WSI has two stages, i) sampling important
patches and modeling them into a fully-connected graph, and ii) converting the
fully-connected graph into a vector representation for classification or
regression purposes. These two stages can be learned end-to-end in a single
training loop. The major novelty of our method is the learning of the
adjacency matrix that defines the connections within nodes. The overall proposed method is
shown in \autoref{fig:wsi_graph} and \autoref{fig:my_label}. The method can
be summarized as follows.

\begin{enumerate}
\item The important patches are sampled from a WSI using a color-based method
  described in~\cite{kalra2019yottixel}. A pre-trained CNN is used to extract
  features from all the sampled patches.
\item The given WSI is then modeled as a fully-connected graph. Each node is
  connected to every other node based on the adjacency matrix. The adjacency
  matrix is learned end-to-end using Adjacency Learning Layer.
\item The graph is then passed through a Graph Convolution Network followed by a
  graph pooling layer to produce the final vector representation for the given
  WSI.
\end{enumerate}

The main advantage of the method is that it processes entire WSIs. The final
vector representation of a WSI can be used for various tasks---classification
(prediction cancer type), search (KNN search), or regression (tumor grading,
survival prediction) and others.\\

\noindent \textbf{Patch Selection and Feature Extraction.} We used the method for patch selection proposed in~\cite{kalra2019yottixel}.  Every WSI contains a
bright background that generally contains irrelevant (non-tissue) pixel
information. We removed non-tissue regions using color thresholds. Segmented
tissue is then divided into patches. All patches are grouped into a pre-set
number of categories (classes) via a clustering method. A portion of all
clustered patches (e.g., 10\%) are randomly selected within each class. Each
patch obtained after patch selection is fed into a pre-trained DenseNet~\cite{huang2017densely} for
feature extraction. We further feed these features to trainable fully connected
layers and obtain final feature vectors each of dimension 1024 representing
patches.\\

\noindent \textbf{Graph Representation of WSI.} We propose a novel method for
learning WSI representation using GCNNs. Each WSI is converted to a fully-connected
graph, which has the following two components.

\begin{enumerate}
\item Nodes $V$: Each patch feature vector represents a node in the graph. 
  The feature for each node is the same as the feature extracted for the corresponding patch.
\item Adjacency Matrix $\mathbf{A}$: Patch features are used to learn the
  $\mathbf{A}$ via adjacency learning layer.
\end{enumerate}

\noindent \textbf{Adjacency Learning Layer.} Connections between nodes $V$ are
expressed in the form of the adjacency matrix $\mathbf{A}$. Our model learns the
adjacency matrix in an end-to-end fashion in contrast to the method proposed in~\cite{tu2019multiple} that
thresholds the $\ell_2$ distance on pre-computed features. Our proposed method also
uses global information about the patches while calculating the adjacency
matrix. The idea behind using the global context is that connection between two
same nodes/patches can differ for different WSIs; therefore, elements in the
adjacency matrix should depend not only on the relation between two patches but
also on the global context of all the patches.

\begin{enumerate}
\item Let $W$ be a WSI and $w_1,w_2,\dots w_n$ be its patches. Each patch $w_i$ is passed through a feature extraction layer to obtain
  corresponding feature representation $x_i$.
\item We use the theorem by Zaheer et al.~\cite{zaheer2017deep} to obtain the
  global context from the features $x_i$. Feature vectors from all patches in
  the given WSI are pooled using a pooling function $\phi$ to get the global
  context vector $c$. Mathematically,
  \begin{equation} c= \operatorname{\phi}(x_1,x_2, \dots, x_n).
  \end{equation} Zaheer et al. showed that such a function can be used as
  an universal set approximator.
\item The global context vector $c$ is then concatenated to each feature vector
  $x_i$ to obtain concatenated feature vector $x_i^\prime$ which is passed through
  MLP layers to obtain new feature vector ${x_i^*} \cdot {x_i^*}$ are the new features
  that contain information about the patch as well as the global context.
\item Features $x_i^*$ are stacked together to form a feature matrix
  $\mathbf{X^*}$ and passed through a cross-correlation layer to obtain adjacency
  matrix denoted by $\underset{n\times n}{\mathbf{A}}$ where each element $a_{ij}$
  in $\mathbf{A}$ shows the degree of correlation between the patches $w_i$ and
  $w_j$. We use $a_{ij}$ to represent the edge weights between different nodes in
  the fully connected graph representation of a given WSI.
\end{enumerate}

\noindent \textbf{Graph Convolution Layers.} Once we implemented the graph
representation of the WSI, we  experimented with two types of GCNN: ChebNets and
GraphSAGE Convolution, which are spectral and spatial methods, respectively.
Each hidden layer in GCNN models the interaction between nodes and
transforms the feature into another feature space. Finally, we have a graph
pooling layer that transforms node features into a single vector
representation. Thus, a WSI can now be represented by a condensed vector, which
can be further used to do other tasks such as classification, image retrieval,
etc.\\

\noindent \textbf{General MIL Framework.} Our proposed method can be used in
any MIL framework. The general algorithm for solving MIL problems
is as follows:

\begin{enumerate}
\item Consider each instance as a node and its corresponding feature as the node
  features.
\item The global context of the bag of instances is learned to calculate the
  adjacency matrix $\mathbf{A}$.
\item A fully connected graph is constructed with each instance as a node and
  $a_{ij}$ in $A$ representing the edge weight between $V_i$ and $V_j$.
\item Graph convolution network is used to learn the representation of the
  graph, which is passed through a graph pooling layer to get a single feature
  vector representing the bag of instances.
\item The single feature vector from the graph can be used for classification or
  other learning tasks.
\end{enumerate}

\section{Experiments}
\label{sec:result}
We evaluated the performance of our model on two datasets i) a popular benchmark
dataset for MIL called MUSK1~\cite{Dua:2019}, and ii) 1026 lung slides from TCGA
dataset~\cite{gutman2013cancer}. Our proposed method achieved a state-of-the-art
accuracy of 92.6\% on the MUSK1 dataset. We further used our model to
discriminate between two sub-types of lung cancer---Lung Adenocarcinoma (LUAD)
and Lung Squamous Cell Carcinoma (LUSC). \\

\noindent\textbf{MUSK1 Dataset. } It has 47 positive bags and 45 negative
bags. Instances within a bag are different conformations of a molecule. The task
is to predict whether new molecules will be musks or non-musks. We performed 10
fold cross-validation five times with different random seeds. We compared our
approach with various other works in literature, as reported
in~\autoref{tab:musk}. The miGraph~\cite{zhou2009multi} is based on kernel
learning on graphs converted from the bag of instances. The latter two
algorithms, MI-Net~\cite{wang2018revisiting}, and
Attention-MIL~\cite{ilse2018attention}, are based on DNN and use either pooling
or attention mechanism to derive the bag embedding.\\

\begin{table}[h]
    \centering
    \begin{tabular}{|l|c|}
        \hline
         \textbf{Algorithm}                         & \textbf{Accuracy} \\
         \hline
         mi-Graph~\cite{zhou2009multi}              & 0.889             \\
         MI-Net~\cite{wang2018revisiting}           & 0.887             \\
         MI-Net with DS~\cite{wang2018revisiting}   & 0.894             \\
         Attention-MIL~\cite{ilse2018attention}     & 0.892             \\
      Attention-MIL with gating~\cite{ilse2018attention} & 0.900             \\
      Ming Tu et al.~\cite{tu2019multiple}          & 0.917             \\
         \textbf{Proposed Method}                   & \textbf{0.926}    \\
         \hline
    \end{tabular}
    \caption{Evaluation on MUSK1.}
    \label{tab:musk}
\end{table}

\begin{figure*}[htbp]
  \centering
  \includegraphics[width=0.98\textwidth]{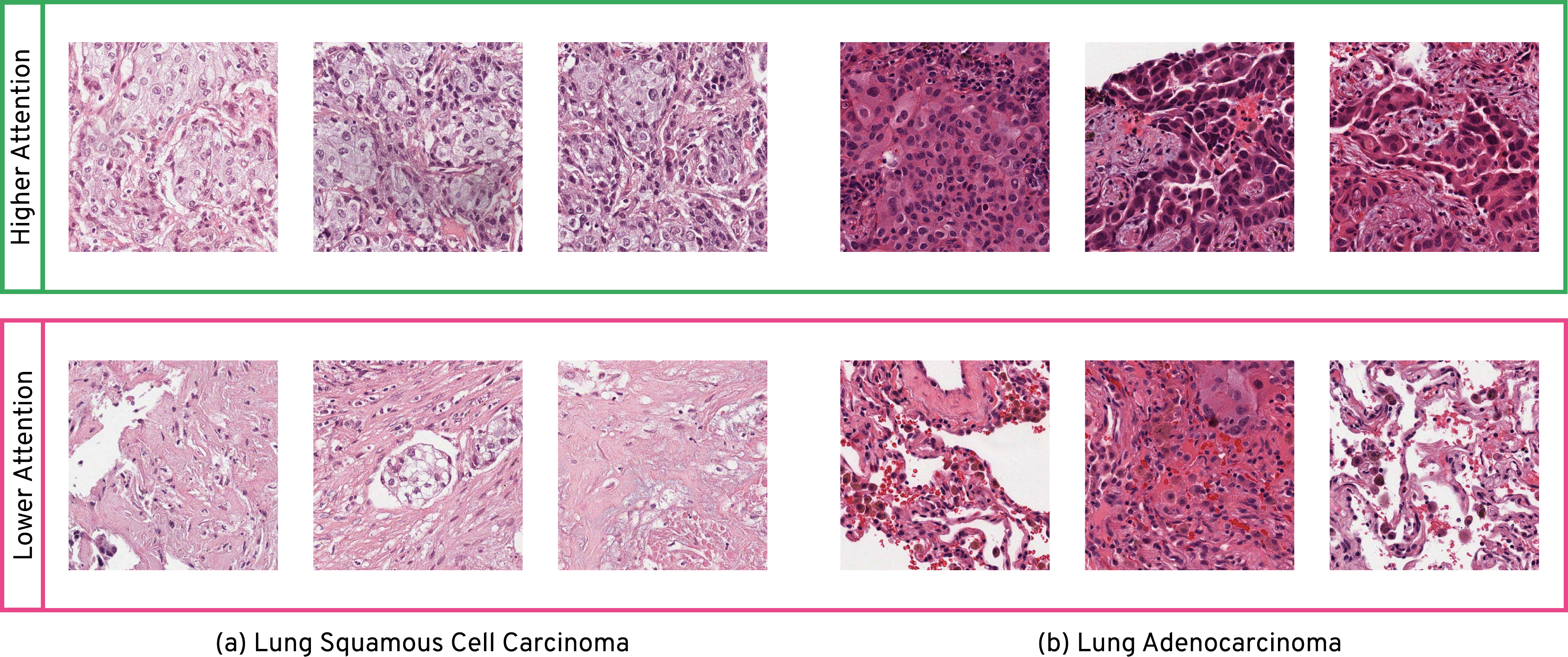}
  \caption{Six patches from two WSIs diagnosed with LUSC and LUAD, 
    respectively. The six patches are selected, such that the first three (top row) are
    highly ``attended'' by the network, whereas the last three (bottom row) least
    attended. The first patch in the upper row is the most attended patch (more
    important) and the first patch in the lower row in the least attended patch
    (less important). }
  \label{fig:attention-patches}
\end{figure*}

\noindent \textbf{LUAD vs LUSC Classification.} Lung Adenocarcinoma (LUAD) and
Lung Squamous Cell Carcinoma (LUSC) are two main sub-types of non-small cell
lung cancer (NSCLC) that account for 65-70\% of all lung
cancers~\cite{graham2018classification}. Automated classification of these two
main subtypes of NSCLC is a crucial step to build computerized decision support
and triage systems.

\begin{figure}[t]
  \centering
  \includegraphics[width=0.5\textwidth]{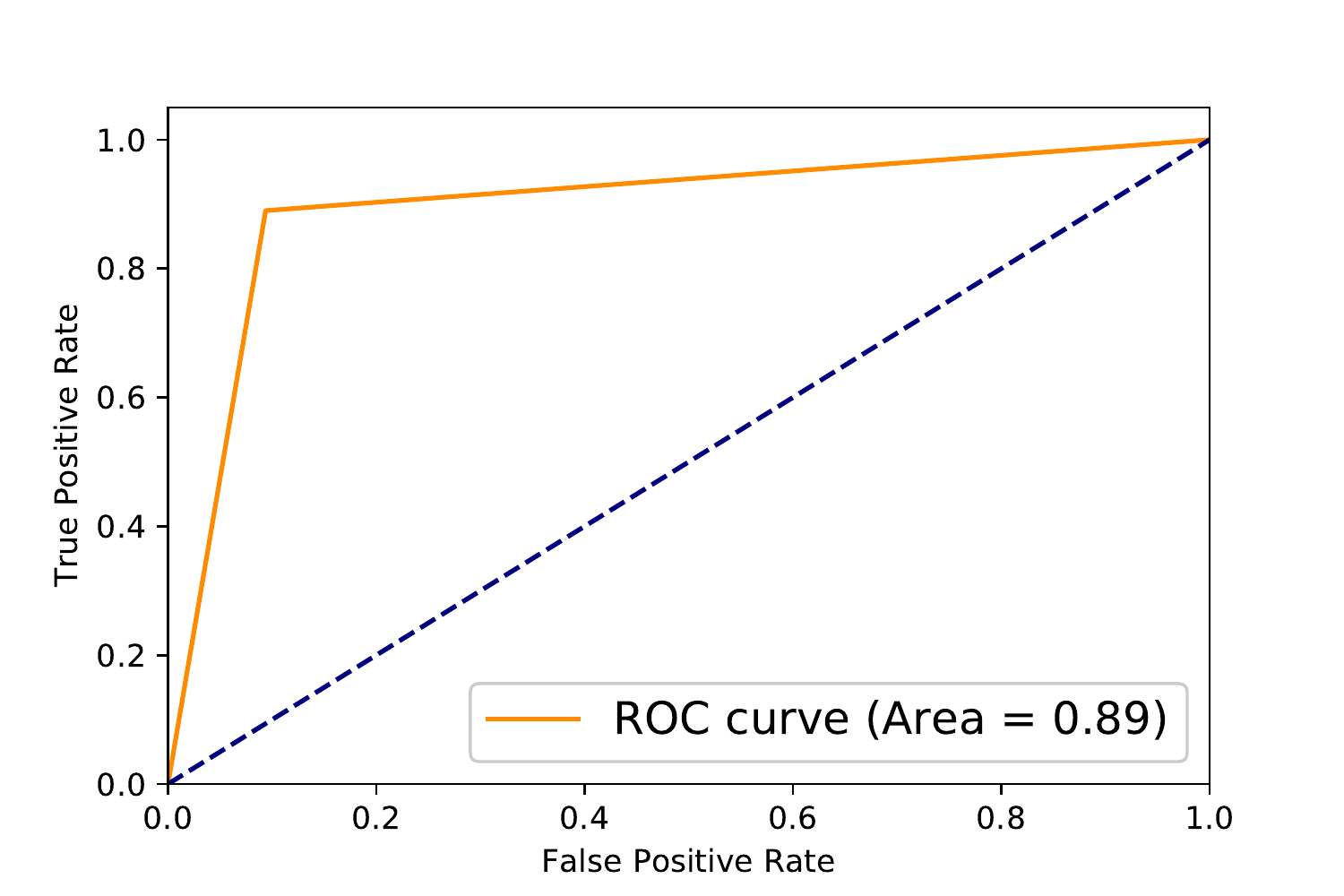}
  \caption{The ROC curve of prediction.}
  \label{fig:roc}
\end{figure}

We obtained 1,026 hematoxylin and eosin (H\&E) stained permanent diagnostic WSIs
from TCGA repository~\cite{gutman2013cancer} encompassing LUAD and LUSC. We
selected relevant patches from each WSI using a color-based patch selection
algorithm described in~\cite{kalra2019yottixel,kalra2020pan}. Furthermore, we extracted image
features from these patches using DenseNet~\cite{huang2017densely}. Now, each
bag in this scenario is a set of features labeled as either LUAD or LUSC. We
trained our model to classify bags as two cancer subtypes. The highest 5-fold
classification AUC score achieved was 0.92, and the average AUC across all folds
was 0.89. We performed cross-validation across different patients, i.e., training
was performed using WSIs from a totally different set of patients than the testing. The results
are reported in~\autoref{tab:our-auc}. We achieved state-of-the-art accuracy
using the transfer learning scheme. In other words, we extracted patch features
from an existing pre-trained network, and the feature extractor was not
re-trained or fine-tuned during the training process. The~\autoref{fig:roc}
shows the receiver operating curve (ROC) for one of the folds. \\

\begin{table}[htb]
  \centering
  \begin{tabular}{|l|c|}
    \hline
    \textbf{Algorithm}                              & \textbf{AUC}  \\
    \hline
    Coudray et al.~\cite{coudray2018classification} & 0.85          \\
    Khosravi et al.~\cite{khosravi2018deep}         & 0.83          \\
    Yu et al.~\cite{yu2016predicting}               & 0.75          \\
    \textbf{Proposed method}                                   & \textbf{0.89} \\
    \hline
  \end{tabular}
  \caption{Performance of various methods for LUAD/LUSC predictions using
    transfer learning. Our results report the average of 5-fold accuracy values.}
  \label{tab:our-auc}
\end{table}

\begin{figure}[t]
  \centering    
  \includegraphics[width=0.45\textwidth]{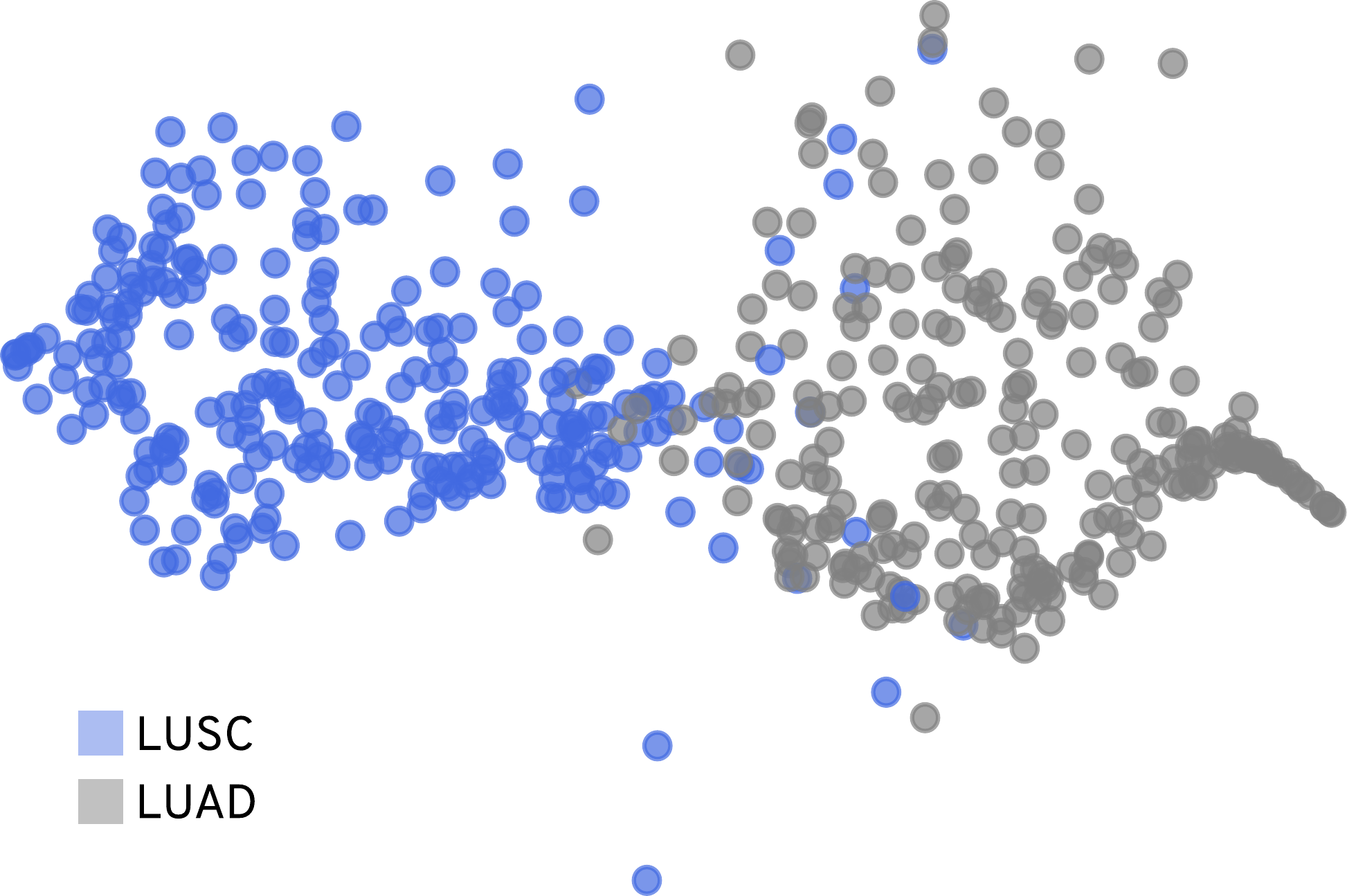}
  \caption{t-SNE visualization of feature vectors extracted after the Graph
    Pooling layer from different WSIs. The two distinct clusters for LUAD and
    LUSC demonstrate the efficacy of the proposed model for disease
    characterization in WSIs. The overlap of two clusters contain WSIs that are
    morphologically and visually similar.}
  \label{fig:tsne_graph}
\end{figure}

\noindent \textbf{Inference.} One of the primary obstacles for real-world
application of deep learning models in computer-aided diagnosis is the black-box
nature of the deep neural networks. Since our proposed architecture uses Global
Attention Pooling~\cite{li2015gated}, we can visualize the importance that our
network gives to each patch for making the final prediction. Such
visualization can provide more insight to pathologists regarding the model's
internal decision making. The global attention pooling layer learns to map patches to
``attention'' values. The higher attention values signify that the model focuses
more on those patches. We visualize the patches with high and low attention
values in~\autoref{fig:attention-patches}.  One of the practical applications of our approach would be for triaging. As new cases are queued for an expert's analysis, the CAD system could highlight the regions of interests and sort the cases based on the diagnostic urgency. We observe that patches
with higher attention values generally contain more nuclei. As morphological features of
nuclei are vitals for making diagnostic decisions~\cite{naik2008automated}, it is interesting to note this property is learned on its own by the network. 
\autoref{fig:tsne_graph} shows the t-SNE plot of features vectors for some of
the WSIs. It shows the clear distinction between the two cancer subtypes, further
favoring the robustness of our method. \\

\noindent \textbf{Implementation Details.} We used PyTorch Geometric library to
implement graph convolution networks~\cite{Fey/Lenssen/2019}. We used
pre-trained DenseNet~\cite{huang2017densely} to extract features from
histopathology patches. We further feed DenseNet features through three
dense layers with dropout ($p=0.5$). \\

\noindent \textbf{Ablation Study. } We tested our method with various different
configurations for the TCGA dataset. We used two layers in Graph Convolution
Network---ChebNet and SAGE Convolution. We found that ChebNet outperforms SAGE
Convolution and also results in better generalization. Furthermore, we
experimented with different numbers of filters in ChebNet, and also different
pooling layers---global attention, mean, max, and sum pooling. We feed the
pooled representation to two fully connected Dense layers to get the final
classification between LUAD and LUSC. All the different permutations of various
parameters result in 32 different configurations, the results for all these
configurations are provided in \autoref{tab:ablation}. It should be noted that
the results reported in the previous sections are based on Cheb-7 with mean
pooling.

\begin{table*}[h]
  \centering
  \begin{tabular}{lrrrr}
\toprule
    configuration   & \textbf{mean}    & attention & max    & add    \\
    \midrule
    \textbf{Cheb-7} & \textbf{0.8889} & 0.8853    & 0.7891 & 0.4929 \\
    Cheb 3\_BN      & 0.8771          & 0.8635    & 0.8471 & 0.5018 \\
    Cheb 5          & 0.8762          & 0.8830    & 0.8750 & 0.5082 \\
    Cheb 3          & 0.8752          & 0.8735    & 0.8702 & 0.5090 \\
    Cheb 5\_BN      & 0.8596          & 0.8542    & 0.7179 & 0.4707 \\
    Cheb 7\_BN      & 0.7239          & 0.6306    & 0.5618 & 0.4930 \\
    SAGE CONV\_BN   & 0.6866          & 0.5848    & 0.6281 & 0.5787 \\
    SAGE CONV       & 0.5784          & 0.6489    & 0.5389 & 0.5690 \\ 
    
\bottomrule
\end{tabular}
\caption{Comparison of different network architecture and pooling method
  (attention, mean, max and sum pooling). \textbf{BN} stands for
  BatchNormalization~\cite{ioffe2015batch}, \textbf{Cheb} stands for Chebnet with
  corresponding filter size and \textbf{SAGE} stands for SAGE Convolution. The
  best performing configuration is Cheb-7 with mean pooling.} 
  \label{tab:ablation}
\end{table*}

\section{Conclusion}
The accelerated adoption of digital pathology is coinciding with and probably
partly attributed to recent progress in AI applications in the field of
pathology. This disruption in the field of pathology offers a historic chance to
find novel solutions for major challenges in diagnostic histopathology and
adjacent fields, including biodiscovery. In this study, we proposed a technique
for representing an entire WSI as a fully-connected graph. We used the graph
convolution networks to extract the features for classifying the lung WSIs into
LUAD or LUSC. The results show the good performance of the proposed approach.
Furthermore, the proposed method is explainable and transparent as we can use
attention values and adjacency matrix to visualize relevant patches.

\balance
{\small \bibliographystyle{ieee_fullname} \bibliography{egbib} }

\end{document}